\documentclass[sigconf]{acmart}
\AtBeginDocument{%
  }

\setcopyright{none}
\copyrightyear{© 2026 The Authors}
\acmDOI{}
\acmISBN{}
\acmConference[Accepted at ICSE-NIER '26]{2026 IEEE/ACM 48th International Conference on Software Engineering}{April 12--18, 2026}{Rio de Janeiro, Brazil}
\acmBooktitle{2026 IEEE/ACM 48th International Conference on Software Engineering (ICSE-NIER '26), April 12--18, 2026, Rio de Janeiro, Brazil}

\usepackage{subcaption}
\usepackage{multirow}

\begin{document}

\title{Context-Adaptive Requirements Defect Prediction through Human-LLM Collaboration}








\author{Max Unterbusch}
\affiliation{%
  \department{paluno -- The Ruhr Institute for Software Technology}
  \institution{University of Duisburg-Essen}
  \city{Essen}
  \country{Germany}}
\email{max.unterbusch@uni-due.de}

\author{Andreas Vogelsang}
\affiliation{%
  \department{paluno -- The Ruhr Institute for Software Technology}
  \institution{University of Duisburg-Essen}
  \city{Essen}
  \country{Germany}}
\email{andreas.vogelsang@uni-due.de}


\begin{abstract}
  Automated requirements assessment traditionally relies on universal patterns as proxies for defectiveness, implemented through rule-based heuristics or machine learning classifiers trained on large annotated datasets.
  However, what constitutes a ``defect'' is inherently context-dependent and varies across projects, domains, and stakeholder interpretations.
  In this paper, we propose a Human-LLM Collaboration (HLC) approach that treats defect prediction as an adaptive process rather than a static classification task.
  HLC leverages LLM Chain-of-Thought reasoning in a feedback loop: users validate predictions alongside their explanations, and these validated examples adaptively guide future predictions through few-shot learning.
  We evaluate this approach using the weak word smell on the QuRE benchmark of 1,266 annotated Mercedes-Benz requirements.
  Our results show that HLC effectively adapts to the provision of validated examples, with rapid performance gains from as few as 20 validated examples.
  Incorporating validated explanations, not just labels, enables HLC to substantially outperform both standard few-shot prompting and fine-tuned BERT models while maintaining high recall.
  These results highlight how the in-context and Chain-of-Thought learning capabilities of LLMs enable adaptive classification approaches that move beyond one-size-fits-all models, creating opportunities for tools that learn continuously from stakeholder feedback.
\end{abstract}



\received{30 September 2025}
\received[accepted]{1 December 2025}

\maketitle

\section{Introduction}

Unnoticed defects, such as ambiguity in natural language requirements, can surface as costly problems in downstream
SE tasks and risk project success \cite{fernandez_naming_2017}.
Addressing this risk, the Requirements Engineering (RE) community has developed automated methods for detecting requirements smells using universal patterns as proxies for defectiveness.
These methods hinge on rule-based approaches \cite{femmerRapidQualityAssurance2017, veizagaAutomatedSmellDetection2024, ferrariDetectingRequirementsDefects2018} requiring complex, handcrafted heuristics, or ML/DL-based approaches \cite{habibDetectingRequirementsSmells2021, ezziniAutomatedHandlingAnaphoric2022, yangAutomaticDetectionNocuous2010} which depend on annotated, sufficiently large datasets --- which are scarce \cite{frattiniLiveExtensibleOntology2022}.
Both approaches are costly to develop and inflexible as they are bound to specific types of defects and natural languages.
Aside from these practical limitations, existing approaches are also conceptually flawed from the perspective of requirements quality as quality-in-use:
What constitutes ``good'' or ``bad'' requirements depends on how well they support downstream SE activities \cite{femmerRequirementsQualityQuality2019}, which is context-dependent.

Our idea is to improve existing smell detection by adding a contextualized defect prediction layer that leverages Human-LLM collaboration through a combination of in-context learning and Chain-of-Thought (CoT) reasoning in an adaptive feedback loop.
The LLM generates reasoning sentences for each defect prediction, which users can accept or reject alongside the prediction itself, creating a growing pool of validated examples with their associated rationales.
Through similarity-based shot selection, the most relevant past examples guide future predictions, enabling the system to continuously adapt to a given development context.
Due to LLMs' pre-trained knowledge, this process could start with zero-shot learning when no examples are available.

To evaluate this approach, we conducted an initial empirical investigation on the case of the weak word smell via the recently released \textsc{QuRE} benchmark \cite{femmerDescriptionComparativeAnalysis2025}.
The dataset contains requirements from automotive manufacturer Mercedes-Benz with weak words, annotated as defect or no defect by internal testing engineers.
The classification of weak word defectiveness is challenging because it requires semantic understanding (see Table~\ref{tab:exampleRequirementsExplanations} for examples).
Simulating our approach on a growing shot pool, we investigated whether LLMs can adapt defect predictions to user-feedback.

Our emerging results show that the approach effectively adapts to context-specific interpretations of defectiveness, even in severely low-data regimes.
We find that incorporating validated explanations alongside labels is critical: HLC with only 20 examples substantially outperforms both standard few-shot prompting without reasoning and BERT models fine-tuned on 320 examples.

With this paper, we suggest a paradigm shift, moving beyond rigid, one-size-fits-all approaches, toward approaches that explicitly incorporate contextual factors and stakeholder-driven judgments.
We expect this Human-LLM collaboration process to extend to other context-dependent tasks in the SE field, such as code reviews, where ``correctness'' often depends on stakeholder perspective and contextual factors.
The paradigm challenges the prevailing assumption that SE automation tools require large pre-annotated datasets with universally agreed ground truths, instead demonstrating that effective quality assurance can emerge from incremental stakeholder feedback.
This opens avenues for future research on context-adaptive SE tooling that learns and evolves with organizations.

\section{Background \& Related Work}

Research on requirements quality assurance has followed two main directions: (i) the use of controlled languages to prevent defects by design, and (ii) verification methods for unconstrained natural language requirements~\cite{ferrariDetectingRequirementsDefects2018}. 
In the latter area, authors defined patterns as universal proxies for potential issues for downstream SE tasks \cite{femmerRapidQualityAssurance2017}, so-called \textit{requirements smells}, and proposed methods for their automated detection.

Rule-based approaches \cite{femmerRapidQualityAssurance2017, ferrariDetectingRequirementsDefects2018, veizagaAutomatedSmellDetection2024} targeted such smells using POS-tagging, dictionaries, and parsing. 
While basic smells such as passive voice and weak words are simple to implement, they typically over-approximate defectiveness, as in practice, most cases are contextually harmless~\cite{krischMythBadPassive2015}.
More sophisticated smells using more narrowly defined patterns, such as vague pronouns and subjective language, are more complex and difficult to maintain with hand-crafted rules.
Several studies employed ML/DL methods to target more concrete defect types such as anaphoric or coordination ambiguity~\cite{yangAutomaticDetectionNocuous2010, ezziniAutomatedHandlingAnaphoric2022, habibDetectingRequirementsSmells2021}.
These statistical methods can better account for context but they are inflexible and require sufficiently large annotated datasets, which are scarce~\cite{frattiniLiveExtensibleOntology2022}. 

Recent work has begun exploring LLMs for requirements quality assessment, targeting abstract quality dimensions such as unambiguity, consistency, or ISO~29148 characteristics~\cite{krishnaUsingLLMsSoftware2024, lubosLeveragingLLMsQuality2024, mahbubCanGPT4Aid2024}. 
However, these studies generally prompt LLMs to provide holistic judgments, often yielding mixed results~\cite{seifertCanLargeLanguage2024, fantechiApplicationsLinguisticTechniques2003} and offering only minimal guidance for the LLM.
Closest to our work, \citet{bashirRequirementsAmbiguityDetection} experimented with few-shot prompting strategies for ambiguity detection and evaluated how well LLMs can post-hoc explain their predictions to practitioners, finding them effective for providing explanations.
In contrast, our approach generates reasoning sentences \emph{before} the verdict, uses them as explanations to gather user feedback, and feeds them back in the demonstrations to guide future predictions.

\section{Human-LLM Collaboration Approach}

\begin{figure}
    \centering
    \includegraphics[width=1\linewidth]{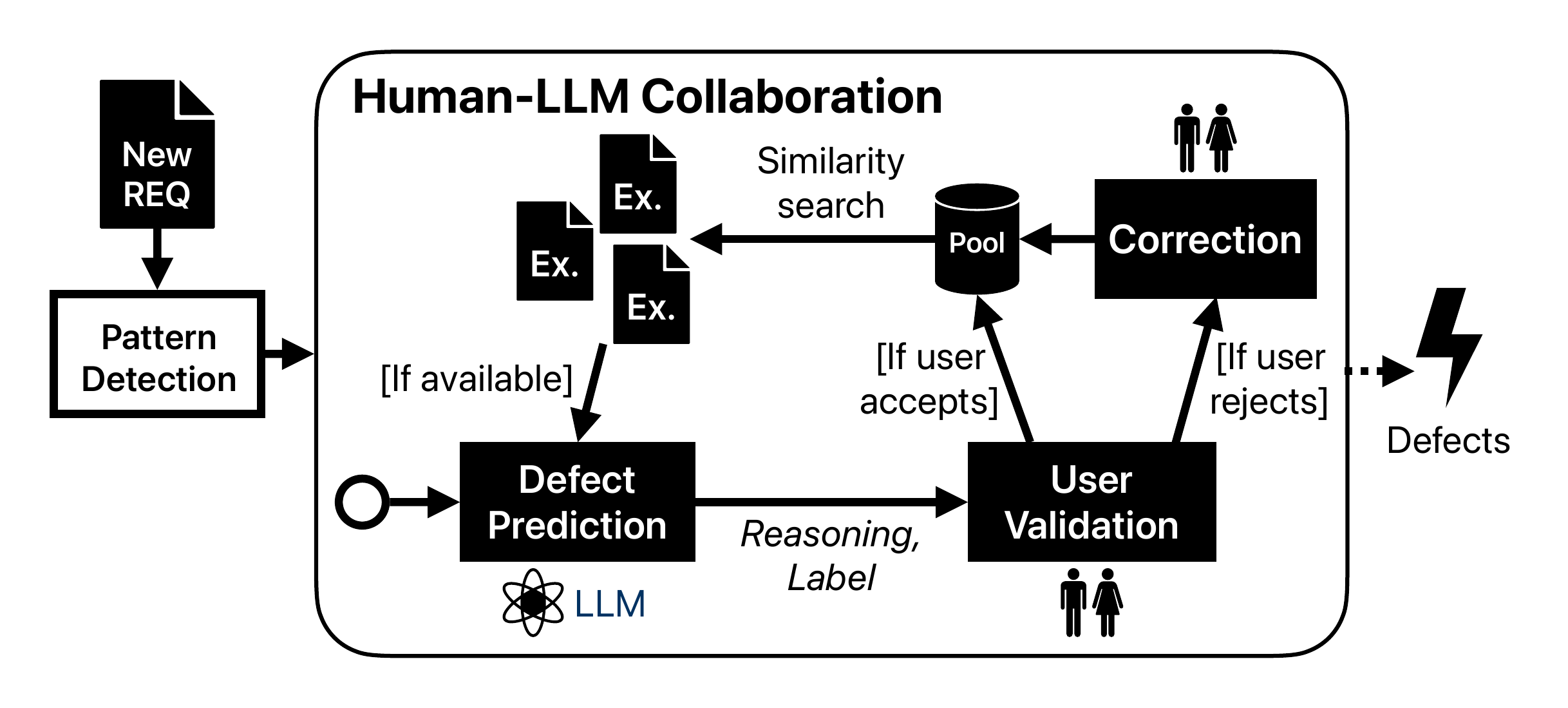}
    \Description{A flow diagram titled “Human–LLM Collaboration.” It shows a cyclical process beginning with a new requirement entering a pattern detection step. The detected pattern is sent to the LLM, which produces a reasoning statement and a defect prediction. The user validation step follows, where the user may accept or correct the reasoning or label. Accepted examples are added to a pool of past examples. When available, similarity search retrieves relevant examples from this pool to provide few-shot context to the LLM for future predictions. The loop continues as the system refines predictions based on validated historical examples.}
    \caption{Human-LLM Collaboration Approach}
    \label{fig:HLC}
\end{figure}

Our Human-LLM collaboration (HLC) approach (Figure~\ref{fig:HLC}) operationalizes defect prediction by building on simple patterns to predict their defectiveness in-context, avoiding the cold start problem entirely.
Starting without any annotated data, we begin with zero-shot CoT prompting.
For an identified pattern (e.g., a weak word), the LLM generates a reasoning sentence before determining defect prediction.
Each finding is explicitly validated by the user, with the CoT reasoning serving as an explanation.
Users may correct the reasoning, the label, or both.
Validated examples (requirement, identified pattern, reasoning, label) are stored in a pool, enabling few-shot prompting.
Using all available examples as shots is impractical due to diminishing returns from redundant examples, recency bias, and context window limitations.
Hence, for every input requirement, shots are retrieved individually via embedding similarity to increase the likelihood of presenting relevant examples.
This allows the LLM to derive its decision from similar historical examples, together with their explicit rationales, to align predictions with stakeholders’ context-specific interpretations of quality.

\textbf{Difference to prior work:}  
Sophisticated defect prediction approaches rely on complex rules or ML/DL methods, which assume universal ground truths and require handcrafted rules or large annotated datasets before deploying static classifiers.
We instead leverage simple, high-recall patterns that traditionally yielded excessive false positives due to their generality.
HLC resolves this by distinguishing defective from benign cases through flexible, context-adaptive Human-LLM collaboration.
Different to standard LLM-based few-shot approaches, HLC builds a shot pool from scratch with validated explanations that guide future predictions, which are usually not available unless crafted manually.

Unlike Active Learning (AL), which starts with annotated data and iteratively queries examples for retraining, our validation loop is the operating mode:
Every prediction is validated, feedback is incorporated via in-context learning, and CoT provides inherent explainability.
Assuming sufficient recall, humans mainly filter benign cases by correcting predictions with minimal effort.

The novelty lies in shifting from pre-annotated datasets to continuously building validated examples with reasoning via Human–LLM collaboration.
We later extend this idea to other SE automation tasks and discuss how tools and process integration can support it.

\section{Study Design}

In this preliminary study, our goal is to estimate how well the HLC approach can perform in a simulated usage scenario.
We draw on the initial case of predicting the defectiveness of weak words to benchmark performance, considering the research question (RQ): \textbf{How effective is HLC in defect predictions of weak words?}

To answer this RQ, we draw on a benchmark of industry requirements from Mercedes-Benz, annotated for weak word defectiveness by their internal testing engineers.
In our experimental setup, we simulate the feedback loop by using the labeled data as stand-ins for user feedback and test increasing pool sizes of validated examples.
As baselines, we compare our HLC approach to (i) an LLM without CoT to assess the performance impact of reasoning, and (ii) fine-tuning of a BERT model to represent the previous status quo of training classifiers in advance.

\subsection{Study Objects}

\textbf{Study Data:}
We draw on the \textsc{QuRE} benchmark~\cite{femmerDescriptionComparativeAnalysis2025}, which comprises 2,111 unique Mercedes-Benz requirements (see Table~\ref{tab:exampleRequirementsExplanations} for examples).
Each requirement contains at least one of 23 weak words from the company's catalog.
The data was annotated for weak word defectiveness by up to three company-internal testers, negotiating labels for difficult cases.
While we have no information about the inter-annotator reliability, this dataset allows us to evaluate how well the approach adapts to QA practices of Mercedes-Benz.

\textbf{Dataset Preparation:}
We de-duplicated requirements appearing with multiple weak words.
If a requirement contains both, a defect and a non-defect weak word, we kept the defective instance since the defective class is the minority.
Since the original distribution of weak words and defects of the dataset is unknown~\cite{femmerDescriptionComparativeAnalysis2025}, we undersampled the non-defective class, yielding a balanced dataset of 1,266 instances (633 per class).

\textbf{Sampling Strategy:}
We designed a nested sampling strategy to track how performance changes as more data is added to the \textit{same} shot pool.
The dataset was randomly split into three folds (422 instances each, stratified by label).
Within each fold, we recursively drew stratified subsets of size 320, 160, 80, 40, and 20, ensuring each smaller pool was contained in the next larger one.
Each pool is evaluated on a separate, cross-assigned fold, so that no instance from a shot pool ever appears in its evaluation set.
A visualization of this sampling strategy is presented in our online material\footnotemark[2].

\textbf{Language Model:}
Since a comparison of LLMs is not at the core of this paper, we conducted all experiments using gpt-4.1-mini (gpt-4.1-mini-2025-04-14).
We chose this model for its cost-efficiency and strong MMLU performance\footnote{\url{https://openai.com/index/gpt-4-1/}}, superior to the best-performing LLMs in prior RE ambiguity studies~\cite{bashirRequirementsAmbiguityDetection}.

\subsection{Approach Implementation}

We define the defect prediction of a weak word as binary classification task, where the model is given an input tuple $X = \langle r,w\rangle$, where $r$ is a natural language requirement and $w$ a weak word contained in $r$, and the objective is to assign a nominal label $y \in Y = \{\text{defect, not defect} \} $.

\textbf{Prompt \& Shot Integration:}
The system prompt (see online material\footnote{\url{https://doi.org/10.6084/m9.figshare.30244633}}) briefly defines weak words and tasks the LLM with deciding whether $w$ makes $r$ ambiguous.
Shots are appended as input-output pairs of requirement + weak word and corresponding reasoning + label.
In the CoT variant, the LLM is explicitly instructed to produce a reasoning sentence before the prediction.
New instances are provided as a user prompt, following the same input structure.

\textbf{Shot Selection:}
We pre-embedded all requirements using the text-embedding-3-small model\footnote{\url{https://platform.openai.com/docs/models/text-embedding-3-small}} by OpenAI.
We experimented with $k \in \{0,12\}$ shots: $k=0$ corresponds to zero-shot (no examples available), while $k=12$ draws the six most similar defect and non-defect examples, respectively, measured by cosine similarity to the target requirement.
To exploit LLM recency bias, shots are ordered so the most similar example appears last~\cite{peysakhovichAttentionSortingCombats2023}.

\textbf{Reasoning Examples:}
To simulate a pool of user-validated reasoning, we generated reasoning sentences for all examples in the pools of size 80 and their subsets.
In line with HLC, explanations were first generated by an LLM; yet unlike in the actual collaboration approach, we conditioned generation on the true label and subsequently vetted each explanation ourselves.
During vetting, we ensured that each rationale was consistent with the label and made small edits where needed.
While in our HLC approach, the reasoning + label are vetted by actual stakeholders, we found plausible reasoning for all 240 requirements (see Table~\ref{tab:exampleRequirementsExplanations} for an extract).

\begin{table*}
\centering
\caption{Example requirements and reasoning explanations (weak words are marked bold)}
\footnotesize
\label{tab:exampleRequirementsExplanations}
\resizebox{\textwidth}{!}{%
\begin{tabular}{@{}r p{0.42\textwidth}p{0.04\textwidth}p{0.45\textwidth}@{}}
\toprule
\textbf{ID} & \textbf{Requirement} & \textbf{Defect} & \textbf{Explanation} \\
\midrule
255 & The TCU is connected to the ORC redundantly via CAN and LIN to execute automatic emergency calls on \textbf{certain} crash levels. & yes & The word 'certain' is used to describe which crash levels trigger automatic emergency calls, yet no specific crash levels are defined, making it unclear which crash levels should trigger emergency calls. \\

92 & In case of a Rear Seat Entertainment System (RSU or Tablet PC), the system shall play the alarm and send it to the \textbf{appropriate} audio output of the selected occupants. & no & The word 'appropriate' refers to the audio output corresponding to the selected occupants, which is contextually clear as it relates to the specific occupant selection mentioned in the requirement. \\

\bottomrule
\end{tabular}
}
\end{table*}

\textbf{Fine-Tuning:}
We compare our LLM-based approaches to fine-tuning a smaller encoder-only model (BERT-base-cased), which has represented the state-of-the-art for requirements classification~\cite{unterbuschExplanationNeedsApp2023}.
To ensure fairness with LLM approaches that highlight weak words, we marked weak word boundaries with special tokens.
The [CLS] representation was used for binary classification via a feedforward layer and softmax activation. 
We fine-tuned separate models on shot pools $\geq 80$, as smaller pools are too limited to achieve reliable results.
Full details are documented in our online material\footnotemark[2].

\section{Study Results \& Discussion}

\begin{figure}
    \centering
    
    \begin{subfigure}[b]{\columnwidth}
        \centering
        \includegraphics[width=0.78\columnwidth]{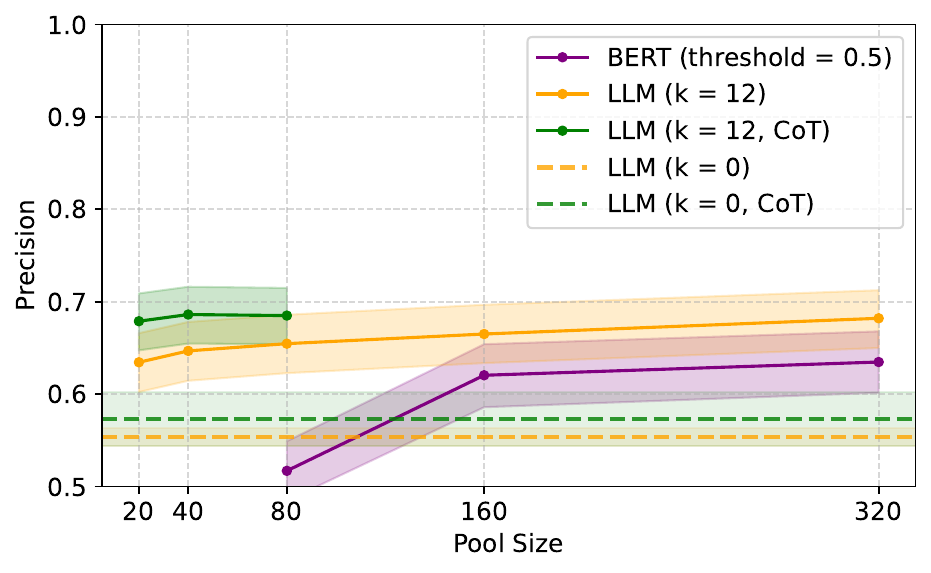}
        \Description{A line chart showing model precision as pool size increases (20, 40, 80, 160, 320). Zero-shot LLM variants start low and stay around 0.55. Few-shot LLM variants (k=12), especially with CoT, reach higher precision around 0.70–0.80. BERT remains consistently lower, roughly between 0.52 and 0.64 across pool sizes.}
        \caption{Precision}
    \end{subfigure}
    
    \begin{subfigure}[b]{\columnwidth}
        \centering
        \includegraphics[width=0.78\columnwidth]{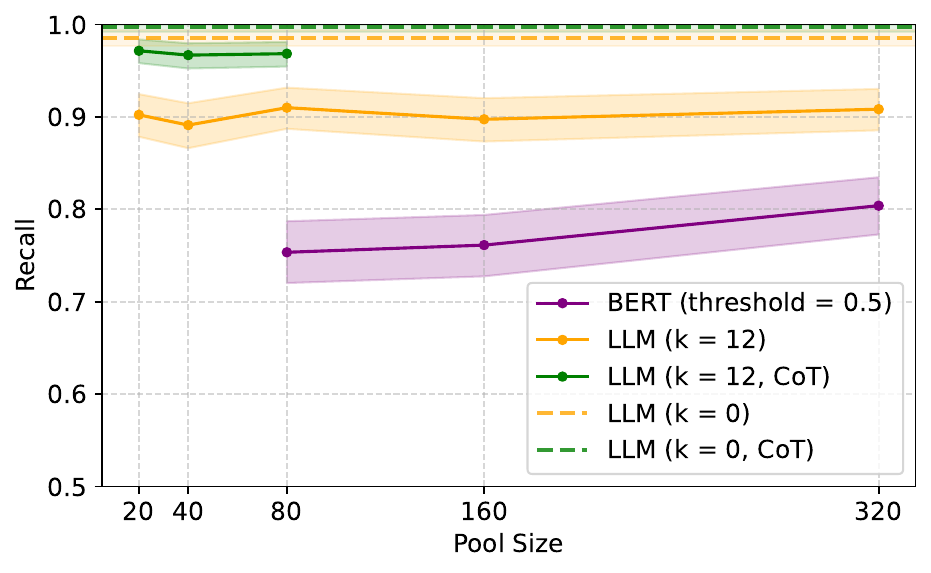}
        \Description{A line chart showing recall across the same pool sizes. Zero-shot LLM variants achieve nearly perfect recall (close to 1.0). Few-shot LLM variants maintain high recall (about 0.90–0.97), with CoT slightly higher. BERT performs substantially worse, around 0.75–0.80 with only small improvements as pool size grows.}
        \caption{Recall}
    \end{subfigure}
    
    \caption{Classification performance across increasing pool sizes (95\% CI obtained by bootstrap resampling, N=10,000)}
    \label{fig:results}
\end{figure}

We evaluate performance using precision, recall, and F1, with defects denoting the positive case.
Because missing a true defect is more costly than flagging a benign case, recall is especially important; however, we report the unweighted F1 since the original dataset distributions are unknown~\cite{femmerDescriptionComparativeAnalysis2025}.
Table~\ref{tab:results} and Figure~\ref{fig:results} summarize the results across all pool sizes.
Confidence intervals were obtained via bootstrap resampling (10,000 iterations) on the concatenated predictions (1,266 per configuration), capturing both within-pool uncertainty and between-pool variability.

\begin{table}
\centering
\caption{Experiment Results}
\footnotesize
\begin{tabular}{@{}r l l r c c c@{}}
\toprule
\textbf{Pool Size} & \textbf{Approach} & \textbf{CoT} & \textbf{$k$} & \textbf{Precision} & \textbf{Recall} & \textbf{F1} \\
\midrule
\multirow{2}{*}{\textbf{--}} & GPT & Yes & 0 & 0.573 & 0.997 & 0.728 \\
 & GPT & No & 0 & 0.553 & 0.986 & 0.709 \\
\addlinespace[3pt]
\multirow{2}{*}{\textbf{20}} & GPT & Yes & 12 & 0.679 & 0.972 & 0.799 \\
 & GPT & No & 12 & 0.634 & 0.902 & 0.745 \\
\addlinespace[3pt]
\multirow{2}{*}{\textbf{40}} & GPT & Yes & 12 & 0.686 & 0.967 & 0.803 \\
 & GPT & No & 12 & 0.647 & 0.891 & 0.750 \\
\addlinespace[3pt]
\multirow{3}{*}{\textbf{80}} & GPT & Yes & 12 & 0.685 & 0.968 & 0.802 \\
 & GPT & No & 12 & 0.655 & 0.910 & 0.761 \\
 & BERT & -- & -- & 0.517 & 0.754 & 0.613 \\
\addlinespace[3pt]
\multirow{2}{*}{\textbf{160}} & GPT & No & 12 & 0.665 & 0.897 & 0.764 \\
 & BERT & -- & -- & 0.620 & 0.761 & 0.684 \\
\addlinespace[3pt]
\multirow{2}{*}{\textbf{320}} & GPT & No & 12 & 0.682 & 0.908 & 0.779 \\
 & BERT & -- & -- & 0.635 & 0.804 & 0.709 \\
\bottomrule
\end{tabular}
\label{tab:results}
\end{table}

Without any validated examples ($k=0$), the LLM-based approaches achieve high recall ($>0.98$), yet only limited precision ($\approx 0.55-0.57$).
CoT reasoning does not meaningfully improve performance in this setting, but it provides explanations --- structurally similar to those we generated beforehand.
These help elicit user feedback, which is an asset for the envisioned collaborative loop.

With only 20 validated examples (10 per class), performance improves markedly.
CoT few-shot prompting ($k=12$) raises precision to $\approx 0.70$ while maintaining high recall ($\approx 0.97$), whereas the non-CoT variant suffers a drop in recall ($\approx 0.90$) alongside lower precision ($\approx 0.63$).

Expanding the shot pool beyond 20 yields only modest precision gains (+4.8 percentage points for the non-CoT variant from 20 to 320 examples), while recall shows no improvement.
This indicates diminishing returns from accumulating larger pools.
In practice, a trade-off arises: forcing continued reasoning corrections improves precision slowly but increases user effort, whereas capping pool growth reduces overhead but leaves more false positives to sort.

Fine-tuning BERT on up to 320 examples results in both lower precision ($\approx 0.64$) and recall ($\approx 0.80$) than the HLC approach with only 20 shots.
Its F1 is even below zero-shot CoT prompting.
While threshold adjustments could bring the practically necessary recall improvement, precision would deteriorate.

\textbf{Answer to RQ.}
HLC effectively adapts defect predictions of weak words as it transitions from zero-shot to few-shot predictions, demonstrating rapid performance gains with only 20 shots.
Incorporating validated explanations in the shots, not just labels, enables HLC to outperform standard prompting approaches that lack such rationales. 
These results should not be overinterpreted as end-to-end performance estimates, because the QuRE benchmark deliberately oversamples challenging cases.

\section{Future Plans}

We aim to advance the HLC paradigm from its current proof-of-concept to a comprehensive framework for context-adaptive SE automation.
Our research agenda spans three complementary directions: tool development and evaluation, extension to broader SE tasks, and refinement of the underlying approach.

Since HLC relies on human-in-the-loop feedback, practical adoption requires tool support that minimizes validation effort.
To this end, we have developed \textit{Requirely}, a prototype tool whose design was informed by our initial findings.
Requirely enables users to flexibly configure checkers beyond weak words, provides context-adaptive requirement defect predictions with explanations, and offers automatically generated improvement suggestions in a rich text editor (watch a demonstration video in our online material\footnotemark[2]).

\textbf{In-Context Evaluation:}
We plan to evaluate tools like Requirely with actual stakeholders in realistic development contexts to assess real-world performance, identify which quality defects can and cannot be effectively detected, and understand the organizational implications of integrating HLC-based tools into SE processes.
We will also investigate user perceptions of HLC compared to static classifiers, examining usability and acceptance.

\textbf{Extension to Other SE Tasks:}
We expect the HLC paradigm to extend beyond requirements quality assurance to other context-dependent classification tasks in SE, where stakeholder-driven judgments may be more favorable than one-size-fits-all solutions.
Specifically, we plan to apply HLC to code review, where code quality is similarly stakeholder-specific and context-dependent.
Similar to requirement smells, code smells could serve as actionable entry points for HLC.
Integration into code review workflows (e.g., GitHub pull requests) could enable HLC to progressively align with reviewer preferences and team standards, speeding up routine quality checks while freeing human experts for other tasks.

\textbf{Approach Refinement:}
Our current study focuses on binary classification with simple pattern-based entry points.
We plan to explore whether HLC can also operate directly on unfiltered artifacts.
We will also investigate extending HLC from classification to generation tasks, such as producing improvement suggestions for requirements or code.
By feeding actually applied improvements back into the shot pool, the system could learn to generate increasingly helpful and contextually appropriate suggestions.
Additionally, we will examine methods to tune precision-recall tradeoffs.

\begin{acks}
Funded by the Deutsche Forschungsgemeinschaft (DFG, German Research Foundation) – Project number: 566352773.
We thank \mbox{Alexander} Korn, Mersedeh Sadeghi, and Henning Femmer for their helpful comments.
\end{acks}

\bibliographystyle{ACM-Reference-Format}
\bibliography{references}

\end{document}